\documentclass[fleqn,usenatbib,useAMS]{mnras}
\usepackage{graphicx}
\usepackage{amsmath}
\usepackage{amssymb}
\usepackage{bm}
\usepackage{color}
\newcommand{\raa}{RAA}

\title[Torus models with LAMOST data]{Torus Models of the Outer Disc of the Milky Way using LAMOST Survey Data}

\author[Q. Wang et al.]
{Qiao Wang$^{1}$\thanks{E-mail:qwang@nao.cas.cn}, Yougang Wang$^{1}$, Chao Liu$^{2}$, Shude Mao$^{3,1,4}$, R. J. Long$^{5,4}$ \\
$^1$Key Laboratory of Computational Astrophysics, National Astronomical Observatories, Chinese Academy of Sciences, Beijing, 100012 China\\
$^2$Key Laboratory of Optical Astronomy, National Astronomical Observatories, Chinese Academy of Sciences, Beijing, 100012 China\\
$^3$Department of Physics, and Center for Astrophysics,  Tsinghua University, 100086 Beijing, China\\
$^4$Jodrell Bank Centre for Astrophysics, School of Physics and Astronomy, The University of Manchester, Oxford Road, Manchester M13 9PL, UK\\
$^5$National Astronomical Observatories, Chinese Academy of Sciences, Beijing, 100012 China\\
}

\begin{document} 
\label{firstpage}

\date{Accepted XXX. Received YYY; in original form ZZZ}
\pagerange{\pageref{firstpage}--\pageref{lastpage}} \pubyear{2016}
\maketitle

\begin{abstract}
With a sample of 48,161 K giant stars selected from the LAMOST DR 2 catalogue, we construct torus models in a large volume extending, for the first time, from the solar vicinity to a Galactocentric distance of $\sim 20$ kpc, reaching the outskirts of the Galactic disc. We show that the kinematics of the K giant stars match conventional models, e.g. as created by Binney in 2012, in the Solar vicinity. However such two-disc models fail if they are extended to the outer regions, even if an additional disc component is utilised. If we loosen constraints  in the Sun's vicinity, we find that an effective thick disc model could explain the anti-centre of the MW.  The LAMOST data imply that the sizes of the Galactic discs are much larger, and that the outer disc is much thicker, than previously thought, or alternatively that the outer structure is not a conventional disc at all. However, the velocity dispersion $\sigma_{0z}$ of the kinematically thick disc in the best-fitting model is about 80 km s$^{-1}$ and has a scale parameter $R_{\sigma}$ for an exponential distribution function of $\sim 19$ kpc.  Such a height $\sigma_{0z}$ is strongly rejected by current measurements in the solar neighbourhood, and thus a model beyond quasi-thermal, two or three thin or thick discs is required.
\end{abstract} 


\section{Introduction}
It is of great importance to construct dynamical models for the Milky Way (MW) to help us understand the structure and formation of the MW and by implication other spiral galaxies. Compared with such galaxies, the MW has the most extensive observational data. Many stellar kinematic surveys with different magnitude limitations and different observing directions have been carried out, such as the Bulge Radial Velocity Assay ~\citep[BRAVA,][]{2007ApJ...658L..29R,2012AJ....143...57K}, the Abundances and Radial velocity Galactic Origins Survey ~\citep[ARGOS,][]{2013MNRAS.428.3660F}, the Apache Point Observatory Galactic Evolution Experiment ~\citep[APOGEE,][]{2008AN....329.1018A,2013ApJ...777L..13M}, SDSS~\citep{2000AJ....120.1579Y}, the Geneva-Copenhagen Survey~\citep[GCS,][]{2004A&A...418..989N}, RAdial Velocity Experiment~\citep[RAVE,][]{2006AJ....132.1645S}, the LAMOST Experiment for Galactic Understanding and Exploration~\citep[LEGUE,][]{2012RAA....12..735D}, the Global Exploration Strategy~\citep[GES,][]{2012Msngr.147...25G} and the GALactic Archeology with Hermes survey~\citep[GALAH,][]{2015MNRAS.449.2604D} etc.

Several methods have been applied to construct dynamical models of the MW. Firstly, Jeans modelling is based on velocity moments of the Jeans equation and connects density, potential and kinematics~\citep{1915MNRAS..76...70J,1919pcsd.book.....J}. For example, ~\citet{2008ApJ...684.1143X,2015ApJ...809..144X}, ~\citet{2014ApJ...794...59K} and \citet{2016MNRAS.463.2623H} employed Jeans models to estimate the mass of the MW using SDSS data. Secondly,  Schwarzschild's orbit-superposition technique ~\citep{1979ApJ...232..236S} relies on replacing the distribution function by a combination of orbit weights and representative orbits. \citet{2012MNRAS.427.1429W,2013MNRAS.435.3437W} have constructed Schwarzschild models of the Galactic bar using BRAVA data. Thirdly, the Made-to-Measure (M2M) method which is particle based ~\citep{1996MNRAS.282..223S} finds the best fit to observations by modifying its particle weights to match the input density and kinematic observables. M2M models of the MW have been constructed by ~\citet{2013MNRAS.428.3478L} (bar) and ~\citet{2015MNRAS.448..713P} (bulge). 

More recently, there has been an increasing interest in torus and action-based distribution function methods for modelling the Milky Way~\citep{2016MNRAS.457.2107S}.  In these methods, stellar distribution functions are constructed from Hamiltonian mechanics action variables \citep[see][]{2008gady.book.....B}. One of the action methods is referred to as torus modelling ~\citep{2008MNRAS.390..429M}. It replaces the orbits in Schwarzschild method by tori and constructs torus libraries. There are other methods. For example, ~\citet{2012MNRAS.426.1328B} connected data from the Geneva-Copenhagen survey of the solar neighbourhood to distribution functions by the St{\"a}ckel fudge method~\citep{2015ApJS..216...29B,2015MNRAS.447.2479S}, and showed in ~\citet{2014MNRAS.439.1231B} that their results are consistent with those from the RAVE survey. Their findings also suggest that data extending significantly beyond the solar radius (as our LAMOST data do) are needed in order to better constrain their models. 

\citet{2014MNRAS.445.3133P} and ~\citet{2015MNRAS.454.3653B} extended earlier modelling and developed the DF of the halo, constraining its parameters by using the RAVE survey and SDSS data within $\sim2.5$ kpc around the Sun. The LAMOST survey has observed a few millions of stars along the Galactic anti-centre direction. This allows us to study the outer disc with many more stars than in previous surveys. In particular, given a limiting magnitude of $r$=17.8 mag, LAMOST K giant stars can reach as far as 90 kpc from the Galactic centre \citep{2014ApJ...790..110L}. The LAMOST sample is thus the best data currently available for the study of the stellar distribution function of the Galactic disc which may extend to about 20 kpc from the Galactic centre.

In this paper, we will construct torus models of the MW using LAMOST data. Our aims are two-fold :
\begin{enumerate}
\item to corroborate or otherwise with LAMOST data results from earlier torus models;
\item to examine whether and how LAMOST data can be used to constrain better the structure of the MW disc.
\end{enumerate} 

The paper is structured as follows. Section 2 describes the LAMOST data. In section 3, we discuss torus modelling, and the algorithms and parameters we use. Results are shown in section 4 and are discussed in section 5.

\section{Data sample}
The Large Area Multi-object Spectroscopic Telescope (LAMOST; also called the Guoshoujing telescope) is a 4-metre reflective Schmidt telescope with a 5-degree field-of-view. 4000 fibres are configured on the wide focal plane, allowing one to observe a few thousands objects simultaneously \citep{2012RAA....12.1197C, 2012RAA....12..723Z}. The LAMOST MW survey targets a few millions of stars with a limiting magnitude of r=17.8 mag. It covers most of the sky from dec=$-10^\circ$ to 60$^\circ$ \citep{2012RAA....12..735D}. Due to the position of the site, during winter, it can  efficiently observe the Galactic anti-centre \citep{2012RAA....12..772Y}. \citet{2014ApJ...790..110L} successfully identified 280,000 K giant stars in the DR1 data. This was later extended to about 450,000 in the DR2 data \citep{2016arXiv160200303H}. Meanwhile, \citet{2015AJ....150....4C} improved the distance estimation for all stars with stellar parameters, including K giant stars, to an accuracy of about 20\% (see the appendix). K giants were chosen as they are bright and detectable at distances far from the Sun. In addition, K giants are present in multiple stellar populations and contain information on different ages and chemistries. 

The selection effect in the LAMOST survey is rather simple ~\citep{2017arXiv170107831L}. The targeting selection strategy mostly depends on the apparent magnitude rather than the color index. Therefore, it would not introduce any bias in kinematics. Moreover, although the initial target selection may be slightly altered by the observations and data reduction, only the very blue or red objects, which may have extremely low signal-to-noise ratio at one end of the spectra (blue end for red objects and vice versa), are possibly dropped by these processes. For the K giant stars used in this work, because their signal-to-noise ratios are quite balanced throughout the whole wavelength coverage, they should not be significantly affected by any systematic bias  during observations and data processing. Therefore, the sample should not have substantial systematic bias in the color index. Consequently, there should be no systematic bias in ages or metallicities, since they are mainly associated with color index.

\begin{figure*}
\centering
\includegraphics[width=\textwidth]{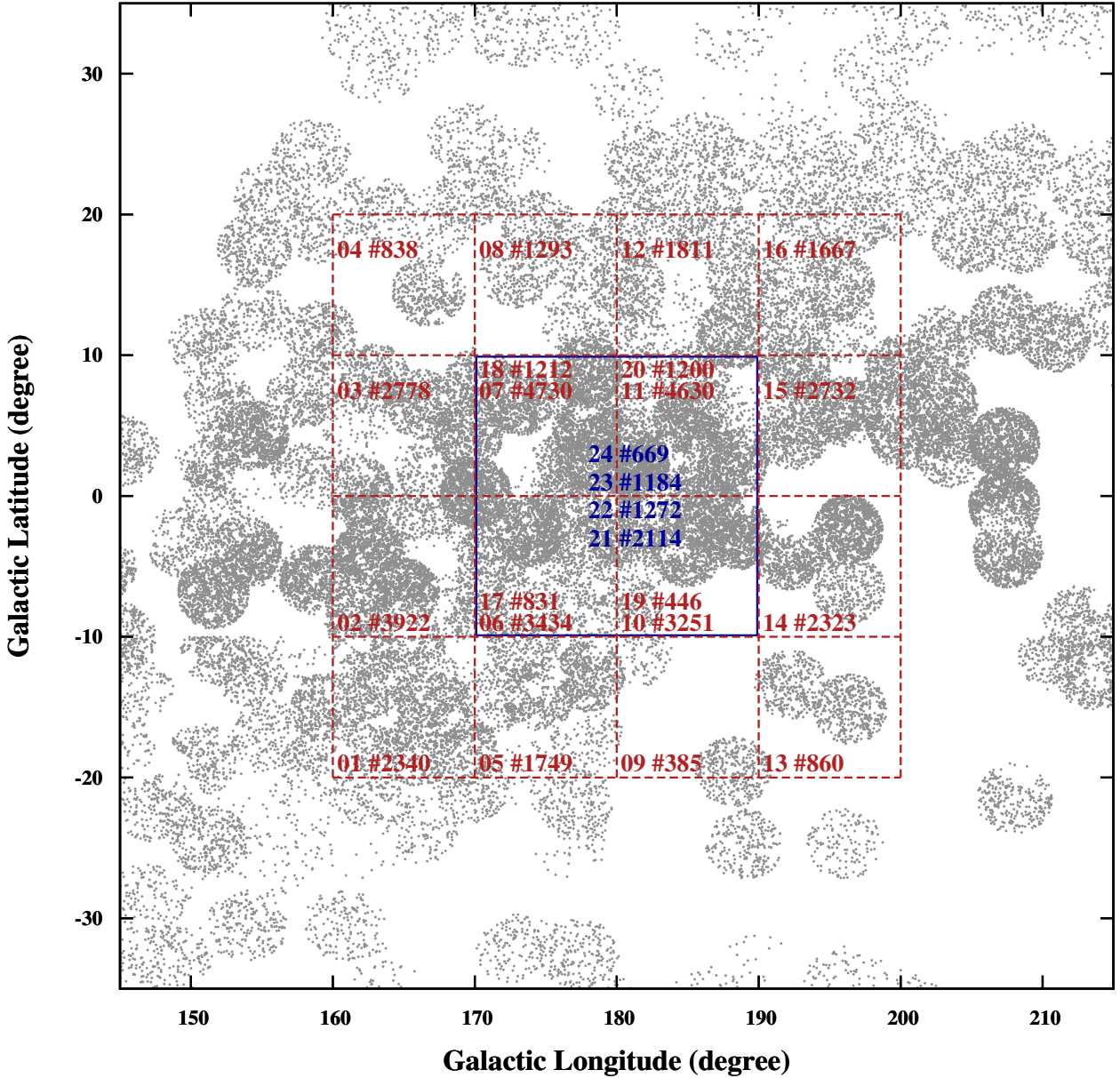}
\caption{The footprint of the LAMOST survey showing the sky blocks in this work (see Table~\ref{tab:block}). The labels denote the identifier of the block and the number of stars in the block (after the \#). Blocks 01 - 16 occupy the nearest 2 kpc subdivision from the Sun, and each block covers 10 $\times$ 10 deg$^2$ sky area. Blocks 17, 18, 19, and 20 cover the same directions as 06, 07, 10, and 11 respectively, but cover the depth of $2 \sim 3$ kpc. Blocks 21, 22 ,23 and 24 cover 20 $\times$ 20 deg$^2$ of sky area of the Galactic anti-centre from $3 \sim 12$ kpc and the blue box is the area covered by blocks 21, 22, 23 and 24. The gray points represent the K giants located within 2 kpc in the survey.}
\label{fig:map}
\end{figure*}

\begin{table}
\centering
\caption{The sky area divided into blocks. There are 26 blocks in total. The first 16 blocks cover the solar neighbourhood. Blocks 17-24 cover $20^\circ \times 20^\circ$ at the anti-centre of the MW out to 12 kpc. The number of K giants is given in the last column of the table.}
\label{tab:block}
\begin{tabular}{lcccr}
\hline
   & longitude range & latitude range  & distance & counts \\
   & ($\deg$)        & ($\deg$)        & (kpc)    &        \\
\hline
01  & (160, 170)  &  (-20, -10) &  [0, 2]    & 2340 \\
02  & (160, 170)  &  (-10, 0)   &  [0, 2]    & 3922 \\
03  & (160, 170)  &  (0, 10)    &  [0, 2]    & 2778 \\
04  & (160, 170)  &  (10, 20)   &  [0, 2]    &  838 \\
05  & (170, 180)  &  (-20, -10) &  [0, 2]    & 1749 \\
06  & (170, 180)  &  (-10, 0)   &  [0, 2]    & 3434 \\
07  & (170, 180)  &  (0, 10)    &  [0, 2]    & 4730 \\
08  & (170, 180)  &  (10, 20)   &  [0, 2]    & 1293 \\
09  & (180, 190)  &  (-20, -10) &  [0, 2]    &  385 \\
10  & (180, 190)  &  (-10, 0)   &  [0, 2]    & 3251 \\
11  & (180, 190)  &  (0, 10)    &  [0, 2]    & 4630 \\
12  & (180, 190)  &  (10, 20)   &  [0, 2]    & 1811 \\
13  & (190, 200)  &  (-20, -10) &  [0, 2]    &  860 \\
14  & (190, 200)  &  (-10, 0)   &  [0, 2]    & 2323 \\
15  & (190, 200)  &  (0, 10)    &  [0, 2]    & 2732 \\
16  & (190, 200)  &  (10, 20)   &  [0, 2]    & 1667 \\
\\
17  & (170, 180)  &  (-10, 0)   &  [2, 3]    & 831  \\
18  & (170, 180)  &  ( 0, 10)   &  [2, 3]    & 1212 \\
19  & (180, 190)  &  (-10, 0)   &  [2, 3]    & 446  \\
20  & (180, 190)  &  (0, 10)    &  [2, 3]    & 1200 \\
\\
21  & (170, 190)  &  (-10, 10)  &  [3, 4]    & 2114 \\
22  & (170, 190)  &  (-10, 10)  &  [4, 5]    & 1272 \\
23  & (170, 190)  &  (-10, 10)  &  [5, 7]    & 1184 \\
24  & (170, 190)  &  (-10, 10)  &  [7, 12]   & 669  \\
\\
25  &  (0, 360)   &  (80, 90)   &  [0, 1.5]  & 270  \\
26  &  (0, 360)   &  (80, 90)   &  [1.5, 3]  & 430  \\
\hline
\end{tabular}
\end{table}

The kinematics of the stellar disc towards the anti-centre are studied in this work. Fig.~\ref{fig:map} shows the footprint  of the survey around the Galactic longitude $l \sim 180^\circ$. Sixteen $10^\circ \times 10^\circ $ sky blocks are distributed around the anti-centre direction, and are numbered from $01$ to $16$. Most contain a few thousand stars, except for blocks $04, 09, 13$. In the area covered by 06, 07, 10, 11, 4 other blocks cover the radial distance from 2 kpc to 3 kpc. They are labelled as 17, 18, 19, 20. Two blocks, 21 and 22, cover the even larger distances of 3-4 kpc and 4-5 kpc, respectively. Block 23 covers 5-7 kpc, and block 24 covers 7-12 kpc. The information is summarized in Table~\ref{tab:block}. The total number of selected K giants in the anti-centre direction is 47,461. In addition, we use blocks 25 and 26 with 700 K giant stars to cover the north Galactic pole to 3 kpc so that the kinematics in the vertical direction may be well detected. The K giants data sample includes sky positions, distances from the Sun, radial velocities and chemical information for all stars.

\begin{figure}
\centering
\includegraphics[width=\columnwidth]{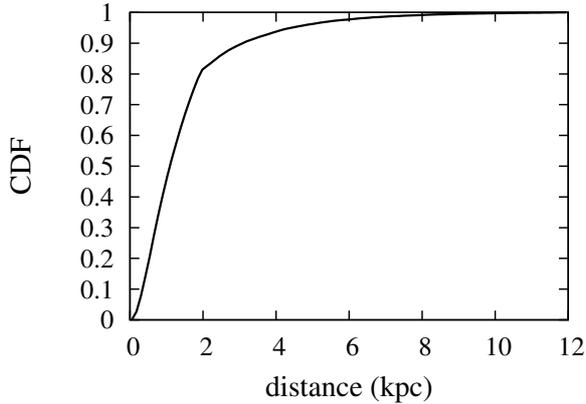}
\caption{The cumulative distribution of K giants as a function of distance from the Sun. 20\% of stars are outside of $\sim$2 kpc.}
\label{fig:df_dist}
\end{figure}

\section{Torus model \& galactic model}

We make no extensions to torus modelling theory nor to the \textit{Torus Mapper} software.  A fuller account of both may be found in \citet{2016MNRAS.456.1982B} and \citet{2016MNRAS.457.2107S}.

\subsection{Action-angle variables}

In a static or quasi-static axisymmetric conservative potential, in addition to the total energy $E$ and the $z$ component of the angle momentum $L_z$, a third integral $I_3$ exists for regular stellar orbits. Since an orbit trajectory projected onto the meridional plane $\{R, z\}$ oscillates about a circular radius, the actions $J_R$ and $J_z$ in phase space \{$ \bm q, \bm p $\}, 
\begin{equation}
J_i = \frac{1}{2 \pi} \oint_{\gamma_i}\bm  p \cdot {\mathbf d} \bm q,
\label{eq:ji}
\end{equation}
replace $E$ and $I_3$ with the third action being $J_\phi=L_z$ ~\citep{1989mame.book.....A,2016MNRAS.457.2107S}. Phase space is now described by the angle action variables \{$\bm \theta,\bm J$\} rather than \{$\bm q, \bm p$\}. The Hamiltonian $H$ does not depend on the angles $\bm \theta$ and so $H \equiv H(\bm J)$. The equations of motion of the stars are given by
\begin{equation}
\dot{\bm J}=0,~~~ \dot{\bm \theta}(t)=\frac{\partial H}{\partial \bm J}=\bm \omega(\bm J).
\end{equation}
It is clear that the angle variables $\bm \theta(t)$ can be written as
\begin{equation}\label{eq:theta}
\bm \theta(t)=\bm \theta(0)+\bm \omega t.
\end{equation}
Orbits in such a system can be expanded in complex variable terms as
\begin{equation}\label{eq:ft}
x(\bm{\theta}, \bm{J})=\sum_{\bf k} X_{\bf k}(\bm J)\exp(i{\bf k}\cdot{\bm \theta}),
\end{equation}
where the sum is over all vectors $\bf k$ with integer components. Combining equation (\ref{eq:ft}) with  equation (\ref{eq:theta}), we find that the spatial coordinates are Fourier series in time, in which every frequency is the linear combination of three fundamental frequencies. In six-dimensional space, an orbit moves only in the three $\bm \theta$ directions, over a surface being topologically equivalent to a three-dimensional torus (3-torus) with the actions $\bm J$ serving to label the orbits.

Only for few specific potentials, such as the St\"ackel and isochrone potentials, can we find the analytic formula for actions in all canonical coordinates. In order to describe the complicated mass model of our Galaxy, a procedure based on canonical transformations is developed for a general potential \citep{1990MNRAS.244..634M, 1993MNRAS.261..584B}. Firstly, a well-understood potential is employed as a reference (\textit{toy}  potential), to connect a point from the configuration velocity space $\{\bm x, \bm v\}$ to action-angle coordinates $\{ \bm \theta^T, \bm J^T\}$, where the superscript $T$ represents \textit{toy}. Then all we need to find is a relationship to connect the general potential to the toy potential. Since a canonical transformation keeps the topological structure, all we need is to find a generating function for the canonical transformation. Such a function is
\begin{equation}
S(\bm \theta^T, \bm J)=\bm \theta^T \cdot \bm J - i \sum_{n \neq 0} S_n(\bm J) e^{in \bm \theta^T},
\end{equation}
in which the series of unknown terms $S_n$~\citep{1990MNRAS.244..634M} is to be determined. Note that $S_n$ are constants for a given $\bm J$. When we gain the approximation of this generating function, $\bm J^T$ can be simply obtained by the equation $\bm J^T = \nabla_{\bm \theta^T}S(\bm \theta^T, \bm J)$. In this way, it is possible to map a point from $\{\bm \theta , \bm J \}$ through $\{\bm \theta^T , \bm J^T \}$, then to the configuration space $\{ \bm x,\bm v \}$. The key point is to compute the coefficients of $S_n$. Instead of direct computation, a fitting method can estimate the $S_n$ by minimizing the error in the Hamilton resulting from the $S_n$ ~\citep{1990MNRAS.244..634M, 1994MNRAS.268.1033K}. 

Such a procedure is complicated. In this work, we use a public package of routines {\it Torus Mapper} \citep{2016MNRAS.456.1982B} to map points from action-angle space to our more usual coordinates (see Section 4). 

\subsection{Mass model and distribution function} 
In this work, we assume that the Milky Way is axisymmetrical with a gas disc, two stellar discs, and a spheroidal halo and bulge. The density of discs is written as ~\citep{1998MNRAS.294..429D}
\begin{equation}
\rho_d(R,z) = \frac{\Sigma_d}{2z_d} \exp \left( -\frac{R}{R_d} -\frac{|z|}{z_d}-\frac{R_h}{R}  \right)
\end{equation}
where $R_d$ is the scale length, $z_d$ is the scale height, and $\Sigma_d$ is the central surface density. The parameter $R_h$ describes a central depression, and is set to be non-zero for the gas disc, and zero for the stellar discs. The spheroidal components have the form
\begin{equation}
\rho_s(R,z) = \frac{\rho_0}{m^\gamma(1+m)^{\beta-\gamma}} \exp \left[ -\left( \frac{ r_0 m}{r_{cut}}  \right)^2 \right]
\end{equation}
where
\begin{equation}
m(R,z)=\sqrt{\left( \frac{R}{r_0} \right)^2 + \left( \frac{z}{q r_0} \right)^2 },
\end{equation}
and $\rho_0$ is the central density, $r_0$ is a scale radius and the parameter $q$ is the axial ratio of the isodensity surfaces. The parameters $\gamma$ and $\beta$ are the slopes for the inner and outer density profiles respectively, and $r_{cut}$ is the cutoff radius. Most of our Milky Way models consist of a gas disc, two stellar discs, a halo and a bulge, except for M11b and M11c (See Table ~\ref{tab:chiq}). It is clear that the potential of each disc component has 4 parameters, while each spheroidal component has 5 parameters ($\rho_0, \gamma, \beta, r_0, r_{cut}$). Therefore, we have 22 parameters in total for one MW density model. The main parameters are listed in Table~\ref{tab:chiq}. 

Following \citet{2012MNRAS.426.1328B}, the distribution function of each single stellar disc can be assumed to be 
\begin{equation}
f(J_R, J_z, L_z) = \frac{\Omega\Sigma\nu}{2\pi^2\sigma_r^2\sigma_z^2\kappa} \exp \left( - \frac{\kappa J_R }{ \sigma_r^2} - \frac{\nu J_z}{\sigma_z^2}  \right) T \left[ \frac{L_z}{L_0} \right], 
\label{eq:df}
\end{equation}
where $\Omega$, $\kappa$, and $\nu$ are the circular, radial, and vertical epicycle frequencies respectively. $\Sigma$ is the radial surface density profile. $T[L_z/L_0]$ is a function of $\left[1+\tanh \left( {L_z}/{L_0} \right) \right]$, with characteristic angular momentum $L_0$. All of them are functions of $L_z$.  In this work, the total distribution function of actions is a combination of the thin and thick discs, their ratio is a free parameter with a default value of 0.7.  

The surface density of a disc is an exponential function
\begin{equation}
\label{eq:sigma}
\Sigma(L_z)=\Sigma_0 \exp \left( -\frac{R_c}{R_d} \right),
 \end{equation}
where radius $R_c$ is derived by assuming a circular orbit with  angular momentum $L_z$.
Given the radius of the Solar circle $R_0$, the vertical and radial velocity dispersions are controlled by the scale parameter $R_{\sigma}$ 
\begin{equation}
\sigma_{r} = \sigma_{r0}\exp\left(  \frac{R_0 - R_c}{R_\sigma} \right), \sigma_{z} = \sigma_{z0}\exp\left(  \frac{R_0 - R_c}{R_\sigma} \right).
\end{equation}

The distribution function of a single disc is controlled by 4 parameters $\sigma_{r0}$, $\sigma_{z0}$, $R_d$, and $R_\sigma$. The $L_0$ truncation scale parameter is fixed at the Torus Mapper value of 9780 kpc~km~s$^{-1}$\citep{2016MNRAS.456.1982B}. An extra parameter is needed to adjust the $ratio$ of the thick to thin discs. In total we have 9 free parameters to control the DF of our two component stellar disc system. 

We build two sets of parameters for the motion of the Sun. Group 1 has $R_0 = 8.0~\rm kpc$, $V_c=220~\rm km ~s^{-1}$, and a Solar motion of (9.58, 10.52, 7.01) km s$^{-1}$ with respect to the local standard of rest (LSR), as described in ~\citet{2015ApJ...809..145T}. Group 2 has $R_0 = 8.5~\rm kpc$, $V_c=244.5~\rm km ~s^{-1}$, and (11.1, 12.24, 7.25) km s$^{-1}$ \citep{2010MNRAS.403.1829S, 2014MNRAS.445.3133P}.

\subsection{Top level modelling process}

There are over 30 parameters in our models. This high number means that it is too expensive computationally to directly constrain all of these parameters simultaneously, even if an MCMC framework is used. As a consequence, we consider the parameters in two groups, those parameters affecting the mass models and those affecting the DFs.  Our modelling procedure to find the best-fitting models and parameters matching our data is a two step procedure using these two groups. We assess the fit of our models to the data using a least $\chi ^2$ approach (see below).

In the first step of our procedure, mass models are examined with fixed DF parameters. Because of the number of mass parameters, 36 mass models are adopted to cover the parameter space. Given an initial set of DF parameters, a best-fitting mass model is found.  For the second step, we then adjust the DF parameters one by one based on the current best-fitting mass model. If the new DF parameters support a different mass model (having a smaller $\chi^2$ value), then this mass model is set as the new reference. We iterate the two-step process until we find the best-fitting parameters for both the mass and DF parts of our model.

The definition of $\chi^2$ we use for a single data block is
\begin{equation}
\chi^2 = \sum_{n} \frac{(p_n^{\rm data} - p_n^{\rm th} ) ^2}{\sigma_n^2},
\label{eq:chi2}
\end{equation}
where $p_n^{\rm data}$ is the line-of-sight velocity distribution with Poisson noise $\sigma_n$ constructed from our LAMOST data, and $p_n^{\rm th}$ is the model prediction. The overall $\chi^2$ value for model comparison purposes is simply the sum of the individual block $\chi ^2$ values. For each block, we compare the model with the data at 40 points. For our 26 data blocks, the total number of points is $n=26 \times 40$, and the degrees of freedom value ($d.o.f$) is $n-32 = 1008$. We ignore any correlation between blocks.

Our data line-of-sight velocity distribution for a block is formed by binning the K giant radial velocity values to give a data distribution histogram.  The model line-of-sight velocity distribution is formed using \textit{Torus Mapper} to create a set of model K giants which are then binned by velocity to give the model distribution histogram.

\textit{Torus Mapper} \citep[TM,][]{2016MNRAS.456.1982B} is an object-oriented C++ toolkit, which provides a user-friendly interface for generating tori. We use its {\it AutoFit} routine to build torus libraries. For each model, a Monte Carlo Markov Chain (MCMC) sampler generates over 2 million points in action space, using our DF parameters and the parameters of our mass model. Every point is taken to represent a model K giant.  Next we use {\it FullMap} to map the stars from action angle variables to configuration space, and the line-of-sight velocity distribution is constructed from the model stars as above.  We have tried modelling with more than 2 million points and rarely find that the modelling accuracy improves, only that modelling becomes more time-consuming. As a guideline, a single, typical model costs roughly 20 cpu hours to produce.

\section{Results}
\label{sec:res}

\subsection{Consistent results in the solar vicinity}
Our current observations provide a constraint for the solar neighbourhood. We check whether our LAMOST data are consistent with the conventional understanding of the solar vicinity, e.g. as in ~\citet{2012MNRAS.426.1328B} from which we use its first potential and corresponding DF parameters as a reference model. Density parameters are listed in Table ~\ref{tab:chiq} and DF parameters can be found found in the first row of Table ~\ref{tab:df}. In what follows, this group of parameter is referred to as B12. It is apparent that LAMOST data are consistent with the B12 model for the nearby blocks 1-16 and 25-26. All of these blocks are located within $\sim$ 2 kpc. This demonstrates that the LAMOST data are consistent with the previous best understanding of kinematics in the solar vicinity. Note that we cannot completely follow the original B12 model due to limitations in the public interface of TM. The DF of B12 can be a superposition of multiple quasi-isothermal components, but we only use fixed thin/thick components because of the TM software interface. We have compared DFs from our procedure and B12, and the amplitude and trends are consistent with each other. 
\begin{table*}
\begin{minipage}{\textwidth}
\centering
\caption{$\chi^2$ for the best-fitting parameters of the distribution functions. The fiducial parameters of the DF are labeled by fid. The row of B12 corresponds to the parameters in \citet{2012MNRAS.426.1328B}. The rows labelled from $1 - 10$ are the models around the second row ($fid$, M11b) . $\chi^2_{\rm inner}$ is estimated by the first 16 blocks in the Solar neighbourhood, $\chi^2_{\rm middle}$ by (17-20), $\chi^2_{\rm outer}$ by (23, 24), $\chi^2_{\rm Pole}$ by (25, 26) and $\chi^2_{\rm Total}$ is estimated by all of 26 blocks (the blocks are defined in Table~\ref{tab:block}).}
\label{tab:df}
\begin{tabular}{@{}r|cccc|cccc|c|ccccc}
\hline
\hline
  & \multicolumn{4}{|c|}{Thin} & \multicolumn{4}{|c|}{Thick}  &ratio & \multicolumn{5}{|c|}{$\chi^2$/$d.o.f$} \\
  & (~ $\sigma_{r0}$ & $\sigma_{z0}$ & $R_d$ &~$R_{\sigma}$~ ) &  ( ~$\sigma_{r0}$ & $\sigma_{z0}$ & $R_d$ & $R_{\sigma}$ ~) & &  \multicolumn{3}{|c|}{( ~~~~~~~~anti-centre ~~~~~~}  &~Pole~ & ~Total~ )\\  
  & km s$^{-1}$ & km s$^{-1}$  & kpc & kpc & km s$^{-1}$ & km s$^{-1}$ & kpc & kpc &  & inner & middle & outer  & & \\ 
\hline 
B12 & 40.1 & 25.6 & 2.58 & 8.93 & 25.8 & 45.0 & 2.11 & 4.04 & 0.772& 3.01 & 4.28 & 32.1 & 1.13 & 4.50 \\
fid & 29.0 & 42.9 & 2.41 & 10.8 & 50.6 & 79.3 & 4.07 & 19.3 & 0.67 & 2.27 & 2.53 & 2.38 & 1.85 & 2.08 \\
1   & 26.1 & 42.9 & 2.41 & 10.8 & 45.5 & 79.3 & 4.07 & 19.3 & 0.67 & 2.67 & 3.54 & 4.85 & 2.26 & 2.61 \\
2   & 31.9 & 42.9 & 2.41 & 10.8 & 55.7 & 79.3 & 4.07 & 19.3 & 0.67 & 3.35 & 2.42 & 5.04 & 3.77 & 2.94 \\
3   & 29.0 & 38.6 & 2.41 & 10.8 & 50.6 & 71.4 & 4.07 & 19.3 & 0.67 & 2.37 & 2.42 & 6.18 & 3.43 & 2.38 \\
4   & 29.0 & 47.2 & 2.41 & 10.8 & 50.6 & 87.2 & 4.07 & 19.3 & 0.67 & 2.32 & 2.50 & 5.36 & 2.17 & 2.25 \\
5   & 29.0 & 42.9 & 2.17 & 10.8 & 50.6 & 79.3 & 3.66 & 19.3 & 0.67 & 2.52 & 2.56 & 3.62 & 3.64 & 2.37 \\
6   & 29.0 & 42.9 & 2.65 & 10.8 & 50.6 & 79.3 & 4.48 & 19.3 & 0.67 & 2.39 & 2.89 & 2.91 & 2.58 & 2.26 \\
7   & 29.0 & 42.9 & 2.41 & 9.72 & 50.6 & 79.3 & 4.07 & 17.4 & 0.67 & 2.21 & 2.88 & 2.81 & 2.50 & 2.15 \\
8   & 29.0 & 42.9 & 2.41 & 11.9 & 50.6 & 79.3 & 4.07 & 21.2 & 0.67 & 2.47 & 2.82 & 4.34 & 3.34 & 2.39 \\
9   & 29.0 & 42.9 & 2.41 & 10.8 & 50.6 & 79.3 & 4.07 & 19.3 & 0.60 & 2.27 & 2.81 & 4.16 & 3.43 & 2.28 \\
10  & 29.0 & 42.9 & 2.41 & 10.8 & 50.6 & 79.3 & 4.07 & 19.3 & 0.74 & 2.39 & 2.37 & 5.96 & 2.42 & 2.32 \\
\hline
\hline
\end{tabular}
\end{minipage}
\end{table*}

\begin{figure*}
\centering
\includegraphics[width=\textwidth]{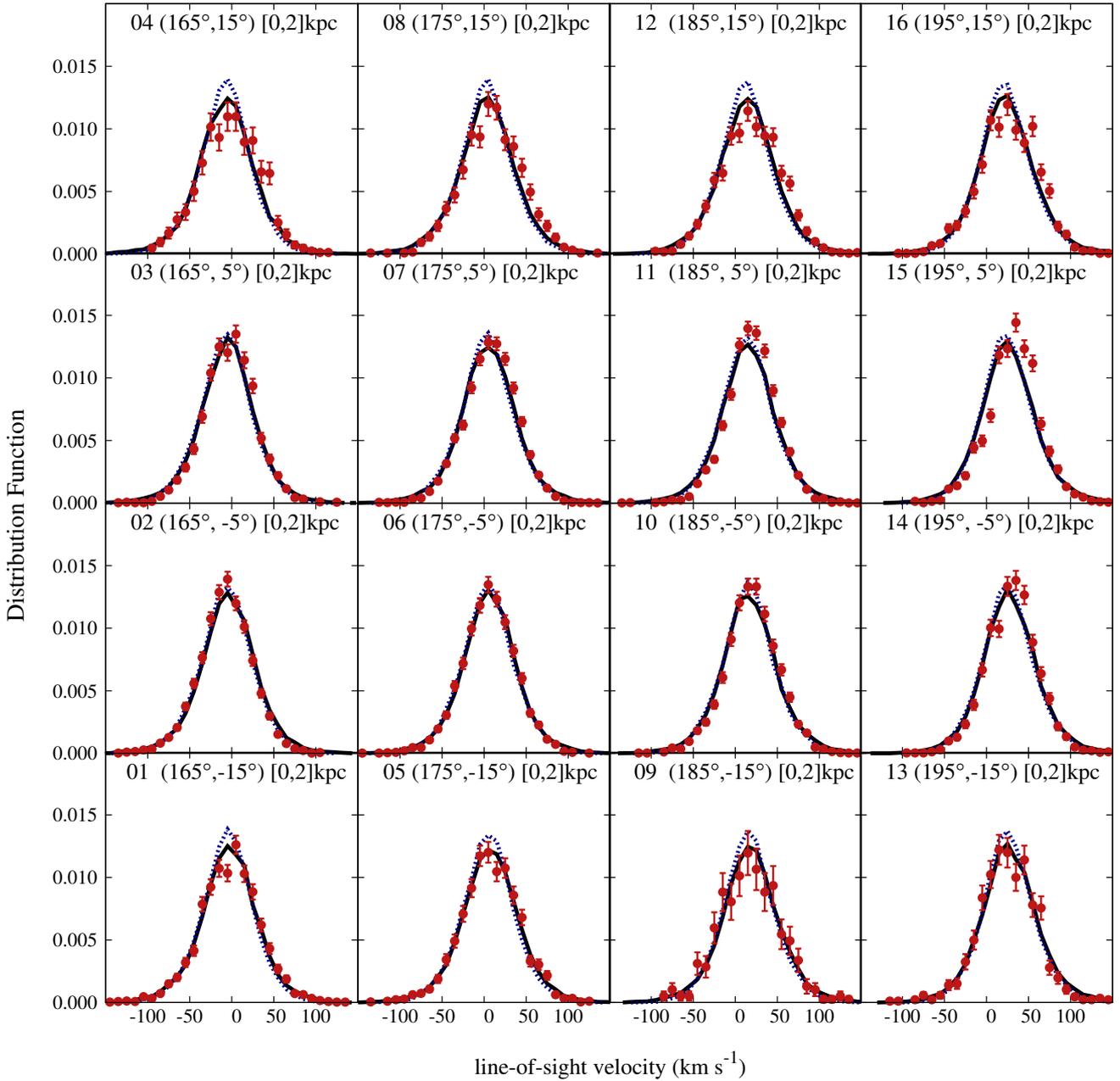}
\caption{The probability distribution function of line-of-sight velocity for blocks $01 - 16$. The upper left number in each panel is the rank of the block. The corresponding distance range is listed in Table~\ref{tab:block}. The black solid and the blue dotted curves represent the predictions from M11b, and B12, respectively. The red points denote the data together with their Poisson error bars. Each block is $10^{\circ} \times 10^{\circ}$. The longitude and latitude of each block centre is indicated together with distance range.}
\label{fig:df16}
\end{figure*}

\begin{figure*}
\centering
\includegraphics[width=\textwidth]{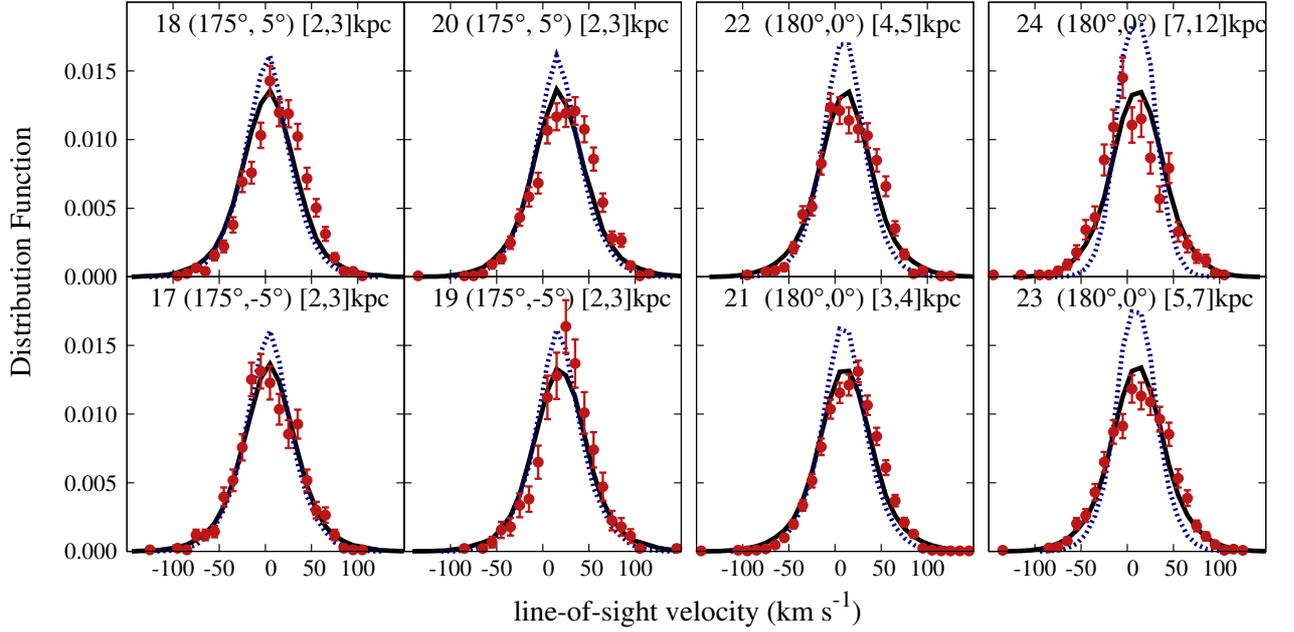}
\caption{The probability distribution function of the line-of-sight velocity for blocks $17 - 24$. The upper left number in each panel is the rank of the block. The corresponding range is listed in Table~\ref{tab:block}. The black solid and the blue dotted curves represent the predictions from M11b and B12, respectively. The red points denote the data together with the Poisson errors. The longitude and latitude of the centre of each block is indicated together with distance range. Blocks $21 - 24$ are $20^{\circ} \times 20^{\circ}$. }
\label{fig:df8}
\end{figure*}

\begin{figure*}
\centering
\includegraphics[width=0.55\textwidth]{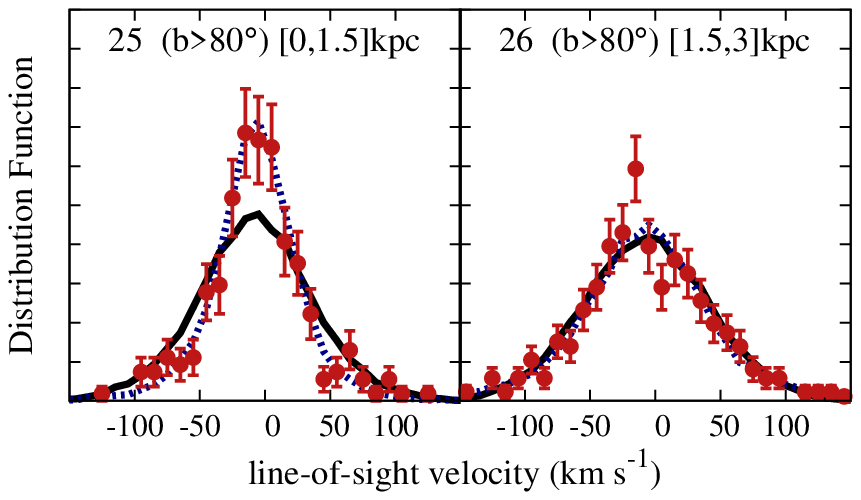}
\caption{The probability distribution density function of the line-of-sight velocity for blocks 25 (left) and 26 (right).  The corresponding range is listed in Table~\ref{tab:block}. The black solid and the blue dotted curves represent the predictions from M11b and B12, respectively. The red points denote the data together with the Poisson errors. The latitude range is indicated together with distance range. The two blocks cover an area of $10^{\circ}\times 10^{\circ}$ in the north Galactic pole direction. }
\label{fig:df2}
\end{figure*}

We illustrate the data l.o.s velocity distribution and our fiducial model. Fig.~\ref{fig:df16} shows that the velocity distributions of block $01 - 16$ are well matched with the models. The results for blocks $17 - 24$, plotted in Fig.~\ref{fig:df8}, correspond to more distant sky regions. The results for blocks 25 and 26, plotted in Fig.~\ref{fig:df2}, correspond to the northern Galactic pole.

Since B12 is constrained only by data within $\sim$ 2 kpc, e.g. GCS, we found that the natural extension of B12 beyond the solar vicinity cannot match the data in the outer region.  As shown in Fig.~\ref{fig:df8}, B12 predicts a relatively small velocity dispersion and the deviation increases as the radius increases. In the next section, we change our parameters to seek a more general model to explain the outer region.

\begin{figure}
\centering
\includegraphics[width=\columnwidth]{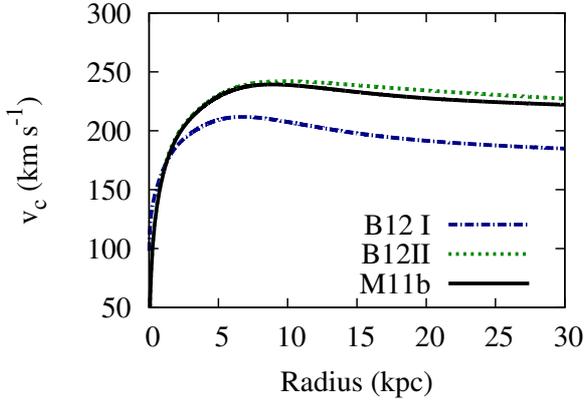}
\caption{Rotation curves for the 3 main mass models. M11b (black solid curve) is the potential of the best-fitting model, and the reference potentials are B12 II (green dot-dashed curve) and
B12 I (blue dotted curve).}
\label{fig:vc}
\end{figure}

\subsection{The hot outer disc}
In order to find our best-fitting model, we experiment with parameters for density and DF across a wide range. Given the thin/thick decomposition, we show the $\chi^2$ values for different models. We start with an initial set of DF parameters based on ~\citet{2012MNRAS.426.1328B} and ~\citet{2014MNRAS.445.3133P}. We change iteratively the parameters and find our fiducial DF parameters, those which give the least $\chi^2$ values within 10\% uncertainty of the current best-fitting model. The DF parameters are listed in the first row of Table~\ref{tab:df}.
Our best-fitting DF parameters are listed in Table ~\ref{tab:df} and labelled as `fid' and with $\chi^2 = 2.08$. This model does describe the hot disc and behaves much better than B12 in the outer disc. However, the velocity dispersion $\sigma_{z0}$ of the thick disc is $\sim 80$ $\rm {km\ s^{-1}}$ and the scale length for velocity dispersion $R_{\sigma}$ is close to 20 kpc. This suggests the 
existence of a hot, extended thick disc reaching much beyond the solar neighborhood. The $R_{\sigma}$ parameter is usually set to be twice the  scale length \citep{1981A&A....95..105V, 1989AJ.....97..139L}. In this work, we set it to be free. \citet{2017arXiv170107831L} found from the radial stellar surface density profile measured with the LAMOST giant stars that the outer disc also extends as far as 19\,kpc in Galactocentric radius. This could be associated with the large $R_{\sigma}$ suggested by the best-fitting model.

Furthermore, we examine the effect with the DF parameters. Around the fiducial parameters, the scale length $R_d$ and dispersion $R_{\sigma}$, velocity dispersion $\sigma_{(r,z)}$ and the ratio between the two discs are considered. We estimate $\chi^2$ for four block ranges. The first range consists of block $\{01 - 16\}$ and roughly corresponds to the range of RAVE data. Its $\chi^2$ value is referred to as $\chi^2_{\rm inner}$. For the middle blocks, $\chi^2_{\rm middle}$ is for blocks $\{17 - 20\}$, and $\chi^2_{\rm outer}$ is for blocks $\{23, 24\}$, which cover the outer region of the MW. 
To verify the robustness of our best-fitting parameters as a fiducial model, ten models are created around the best-fitting point in parameter space. We find that $\chi^2$ increases quickly  when $\sigma_{z0}$  decreases to $ \sim 70$ km $s^{-1}$ (model 3 in Table ~\ref{tab:df}). Thus the fiducial model appears to be robust.

Given the fiducial DF parameters (first row in Table~\ref{tab:df}), we examine 36 mass models of the MW. The notations of P14, B12I, B12II and BT08 represent the parameters from~\citet{2014MNRAS.445.3133P}, ~\citet{2012MNRAS.426.1328B} and ~\citet{2008gady.book.....B}, respectively.  M11b denotes the {\it best} model and M11c denotes the {\it convenient} model from ~\citet{2011MNRAS.414.2446M}. The other mass models are from ~\citet{1998MNRAS.294..429D}. The results are listed in Table~\ref{tab:chiq}. The solid curve in the DF Figure is the best-fitting model. Any deviations become more apparent in Figs.~\ref{fig:df8} and ~\ref{fig:df2}.

\begin{table*}
\begin{minipage}{1.0\textwidth}
\caption{The reduced $\chi^2$ for different mass models. The main parameters of discs, bulge and dark halo are listed in the table. G1 and G2 correspond to the two groups with differing solar position and motions. The best-fitting model is M11b in Group 2. The notation of different mass models is described in Sec.~\ref{sec:res}. }
\centering
\label{tab:chiq}
\begin{tabular}{@{}l|cc|cc|cc|ccc|ccc|cc}
\hline
\hline
& \multicolumn{2}{|c|}{Thin} & \multicolumn{2}{|c|}{Thick} & \multicolumn{2}{|c|}{Gas} & \multicolumn{3}{|c|}{bulge} & \multicolumn{3}{|c|}{halo} &  \multicolumn{2}{|c|}{$\chi^2$/$d.o.f$}  \\
& ( $\Sigma_0$~~& ~$R_d$ ) & ( $\Sigma_0$~~ & ~$R_d$ )& ( $\Sigma_0$~~ & ~$R_d$ )& ( $\rho_0$~ & q  & $r_0$ ) & ( $\rho_0$~ & q & $r_0$ ) & G1 & G2 \\
& M$_{\odot}$/kpc$^2$ & kpc & M$_{\odot}$/kpc$^2$ & kpc & M$_{\odot}$/kpc$^3$ & kpc & M$_{\odot}$/kpc$^3$ &   & kpc & M$_{\odot}$/kpc$^3$  & & & &  \\
\hline
P14   & 5.71e8 & 2.68 & 2.51e8 & 2.68 & 9.45e7 & 5.36 & 9.49e10 & 0.5 & 0.075 & 1.81e7 & 1   & 14.4 & 4.32 & 2.59\\
B12I  & 1.02e9 & 2.4  & 1.14e6 & 2.4  & 7.30e7 & 4.8  & 1.26e9  & 0.8 & 1.09  & 7.56e8 & 0.6 & 1    & 2.53 & 2.88\\
B12II & 7.68e8 & 2.64 & 2.01e8 & 2.97 & 1.16e8 & 5.28 & 9.49e10 & 0.5 & 0.075 & 1.32e7 & 1   & 16.5 & 4.27 & 2.54\\
M11c  & 7.53e8 & 3.0  & 1.82e8 & 3.5  &   -    &   -  & 9.41e10 & 0.5 & 0.075 & 1.25e7 & 1   & 17   & 4.36 & 2.29\\
M11b  & 8.17e8 & 2.9  & 2.09e8 & 3.31 &   -    &   -  & 9.56e10 & 0.5 & 0.075 & 8.46e6 & 1   & 20.2 & 3.73 & 2.08\\
BT08  & 1.18e8 & 2.0  & 1.66e9 & 2.0  & 1.32e8 & 4.0  & 7.11e8  & 0.8 & 3.83  & 4.27e8 & 0.6 & 1    & 2.80 & 2.37\\

1     & 1.65e9 & 2.0  & 1.32e8 & 4.0  & 1.18e8 & 2.0  & 7.11e8  & 0.8 & 3.83  & 4.27e8 & 0.6 & 1    & 3.14 & 2.51 \\
10    & 1.59e9 & 2.0  & 1.27e8 & 4.0  & 1.13e8 & 2.0  & 5.0e8   & 0.8 & 5.0   & 5.0e8  & 0.6 & 1    & 12.1 & 6.52 \\

2     & 1.02e9 & 2.4  & 1.14e8 & 4.8  & 7.30e7 & 2.4  & 1.26e9  & 0.8 & 1.09  & 7.56e8 & 0.6 & 1    & 3.08 & 2.81 \\
20    & 1.03e9 & 2.4  & 1.14e8 & 4.8  & 7.32e7 & 2.4  & 1.0e9   & 0.8 & 5.0   & 6.0e8  & 0.6 & 1    & 41.6 & 28.5 \\
2.1   & 1.05e9 & 2.4  & 1.17e8 & 4.8  & 7.52e7 & 2.4  & 1.06e9  & 0.9 & 1.19  & 7.54e8 & 0.6 & 1    & 3.18 & 2.58 \\
2.2   & 9.82e8 & 2.4  & 1.09e8 & 4.8  & 7.0e7  & 2.4  & 1.58e9  & 0.7 & 1.0   & 7.62e8 & 0.6 & 1    & 3.16 & 2.58 \\
2.3   & 9.19e8 & 2.4  & 1.02e8 & 4.8  & 6.57e7 & 2.4  & 1.70e9  & 0.6 & 1.0   & 7.59e8 & 0.6 & 1    & 3.19 & 2.58 \\
2.4   & 8.27e8 & 2.4  & 9.20e7 & 4.8  & 5.9e7  & 2.4  & 1.84e9  & 0.5 & 1.0   & 7.29e8 & 0.6 & 1    & 2.76 & 2.62 \\
2a    & 1.03e9 & 2.25 & 1.18e8 & 4.5  & 7.36e7 & 2.25 & 9.17e5  & 0.8 & 22.8  & 3.96e8 & 0.6 & 1    & 2.70 & 3.49 \\
2b    & 9.87e8 & 2.55 & 1.07e8 & 5.1  & 7.06e7 & 2.55 & 1.42e9  & 0.8 & 1.73  & 7.87e8 & 0.6 & 1    & 3.60 & 2.21 \\
2c    & 1.03e9 & 2.4  & 1.14e8 & 4.8  & 7.33e7 & 2.4  & 3.05e7  & 0.8 & 6.19  & 5.99e8 & 0.6 & 1    & 3.15 & 2.71 \\
2d    & 1.05e8 & 2.4  & 1.16e8 & 4.8  & 7.48e7 & 2.4  & 6.16e7  & 0.8 & 21.8  & 6.75e8 & 0.6 & 1    & 2.79 & 2.77 \\
2e    & 1.07e9 & 2.4  & 1.09e8 & 4.8  & 7.68e7 & 2.4  & 9.08e8  & 0.8 & 1.80  & 6.58e8 & 0.6 & 1    & 3.08 & 2.37 \\
2f    & 9.40e8 & 2.4  & 1.14e8 & 4.8  & 6.71e7 & 2.4  & 4.17e8  & 0.8 & 1.0   & 3.0e8  & 0.6 & 1    & 3.05 & 2.77 \\
2g    & 1.01e9 & 2.4  & 1.12e8 & 4.8  & 7.21e7 & 2.4  & 8.33e8  & 0.8 & 1.92  & 9.53e8 & 0.6 & 1    & 2.84 & 2.77 \\
2h    & 8.66e8 & 2.4  & 9.63e7 & 4.8  & 6.19e7 & 2.4  & 1.07e9  & 0.8 & 2.88  & 1.23e9 & 0.6 & 1    & 3.06 & 2.28 \\
2i    & 7.85e8 & 2.4  & 8.73e7 & 4.8  & 5.6e7  & 2.4  & 8.51e8  & 0.3 & 1.0   & 3.38e8 & 0.6 & 1    & 2.78 & 3.04 \\
2L    & 1.02e9 & 2.4  & 1.13e8 & 4.8  & 7.3e7  & 2.4  & 1.66e9  & 0.8 & 1.09  & 7.56e8 & 0.6 & 1    & 3.49 & 2.53 \\
2S    & 1.02e9 & 2.4  & 1.13e8 & 4.8  & 7.30e7 & 2.4  & 9.13e8  & 0.8 & 1.09  & 7.56e8 & 0.6 & 1    & 2.84 & 3.37 \\

3     & 6.42e8 & 2.8  & 9.06e7 & 5.6  & 4.59e7 & 2.8  & 1.18e8  & 0.8 & 2.29  & 3.0e8  & 0.6 & 1    & 2.87 & 2.46 \\
4     & 4.32e8 & 3.2  & 7.28e7 & 6.4  & 3.09e7 & 3.2  & 2.66e8  & 0.8 & 1.90  & 3.0e8  & 0.6 & 1    & 3.05 & 2.50 \\
40    & 4.26e8 & 3.2  & 7.18e7 & 6.4  & 3.04e7 & 3.2  & 2.20e8  & 0.8 & 2.77  & 3.0e8  & 0.6 & 1    & 3.26 & 2.18 \\
4.1  & 4.52e8 & 3.2   & 7.63e7 & 6.4  & 3.23e7 & 3.2  & 2.24e8  & 0.9 & 1.93  & 3.0e8  & 0.6 & 1    & 3.01 & 2.48 \\
4.2  & 4.05e8 & 3.2   & 6.84e7 & 6.4  & 2.89e7 & 3.2  & 3.18e8  & 0.7 & 1.89  & 3.0e8  & 0.6 & 1    & 3.09 & 2.28 \\
4.3  & 3.67e8 & 3.2   & 6.20e7 & 6.4  & 2.63e7 & 3.2  & 3.88e8  & 0.6 & 1.91  & 3.0e8  & 0.6 & 1    & 3.06 & 2.31 \\
4.4  & 3.41e8 & 3.2   & 5.74e7 & 6.4  & 2.44e7 & 3.2  & 4.26e8  & 0.5 & 1.99  & 3.0e8  & 0.6 & 1    & 2.95 & 2.40 \\
4a   & 4.22e8 & 3.0   & 7.37e7 & 6.0  & 3.02e7 & 3.0  & 7.56e8  & 0.8 & 1.0   & 3.0e8  & 0.6 & 1    & 2.73 & 2.95 \\
4b   & 4.33e8 & 3.4   & 7.09e7 & 6.8  & 3.09e7 & 3.4  & 5.21e7  & 0.8 & 5.24  & 3.0e8  & 0.6 & 1    & 3.50 & 2.34 \\
4c   & 4.25e8 & 3.2   & 7.17e7 & 6.4  & 3.04e7 & 3.2  & 1.29e9  & 0.8 & 1.0   & 4.15e8 & 0.6 & 1    & 2.90 & 2.36 \\
4d   & 4.12e8 & 3.2   & 6.95e7 & 6.4  & 2.95e7 & 3.2  & 1.10e8  & 0.8 & 5.24  & 6.51e8 & 0.6 & 1    & 2.96 & 2.21 \\
\hline
\hline
\end{tabular}
\end{minipage}
\end{table*}

The $\chi^2$ values of the Group 2 models are in general smaller than Group 1. This implies that the solar motion or position in Group 1 may have been systematically under-estimated. In what follows, we only show the results for Group 2.

Besides the DF parameters, the mass models also favour a large disc. The mass models with larger $R_d$, such as 40, 2b, 2d and 4d et c., tend to have smaller $\chi^2$ values. More sophisticated mass models will be considered in a future work.

\subsection{Tension between the models and data}
From the analysis in the last section, our fiducial parameters can roughly match the DF, but the two wings of the l.o.s velocity DF in blocks 25 and 26 do not show the characteristics of halo stars (See  Fig.~\ref{fig:df8}). On the other hand, B12 can explain the RAVE data at the Galactic pole better than our model (see Fig.~\ref{fig:df2}). Our preferred model gives a larger velocity dispersion than the block 25 data. However, we cannot find a model to match both sides under the assumption of a two component decomposition. An improvement in the theory is required. The simplest way is to add an extra component for the outer region beyond the solar vicinity.

Since the scale radius and velocity dispersion of the thick disc are both larger than the thin disc, the long thick disc is more like an envelope of the short, thin disc in the MW outskirts. Thus we add the third disc based around the fiducial parameters, e.g. $\sigma_z \sim 80$ km$s^{-1}$. Around these initial parameters, we run several dozens of models to find a `better' model. We find that some models can fit (with $\chi^2 \leq 2$) the outer data, at about (15, 20) kpc,  but fit poorly ($\chi^2 \geq 4$) the intermediate region about (12, 15) kpc and vice versa. Because we control the mass fraction of the third component to below 0.3 to hold a relatively small $\sigma_z$ in the Solar vicinity, the intermediate region is not dominated by the third disc and it only has a significant effect in the outermost region.

The thin/thick disc decomposition is well known for the solar vicinity, but the parameters and cause of the thick disc (geometrical, kinematical, and chemical and so on) are unclear. A straightforward idea is that the thick disc is formed from the heating of old stars while young stars are in the thin disc. Therefore the geometrically-thick disc should correspond to the old population. Current observations indicate that there is a radial age gradient ~\citep{2016arXiv160901168M}, corresponding to a complex kinematic structure. However, for discs, a possible flaring structure can also exist in the disc outskirts ~\citep{2013ApJ...779..115B, 2016ApJ...823...30B}. 

The number of K giants in this work is insufficient to tell  which kinds of structures are required. The previous thin/thick decomposition is however rejected as are the modified (our fiducial) 2-disc model, and the 3-disc model. If we introduce more components with different scale lengths and velocity dispersions, perhaps a solution for the whole region can be developed, but it would require a complex form of DF. In summary, our  current torus model still needs to be developed further theoretically.

\section{Discussion and Summary}
We have successfully met our aims as set out in the Introduction. Our results corroborate earlier results and we have produced a model for the Milky Way's outer disc.

It is the first time the torus approach has been applied to the LAMOST survey. The torus mapper package TM \citep{2016MNRAS.456.1982B} has been employed to construct our torus models for different density models with different DF parameters. The data are well matched by the model with the mass model of M11b and by our fiducial DF parameters (the first row in Table~\ref{tab:df}), except for the Solar neighborhood. We are forced to have a thick disc that is much more extended and thicker than previously thought. If such a component exists, more careful analyses of age and metallicity information should yield more solid evidence. Given the quasi-thermal disc assumption, there must be a sharp transition of the thin/thick model from the solar vicinity to the outer region. We also test whether a simple additional disc can match the data.  The answer is in the negative, which suggests that a more complex DF is needed to accommodate the outer l.o.s velocity distribution as revealed by the LAMOST K giant data. 

The advantage of LAMOST data is its range, which allows us to study the stellar distribution over 2 kpc. It has helped to rule out some mass models, such as mass model numbers 10 or 20 (see Table~\ref{tab:chiq}). However the tangential velocities of those distant stars are unconstrained. It is still insufficient to completely break the degeneracy between individual components or parameters within the mass models. Meanwhile, the information on age and metallicity needs to be taken into account in future models. 

Although the spatial distribution of stars in block 9 is extremely inhomogeneous, it does not introduce any apparent bias in the DF.  The  non-uniformity of spatial distribution in a block is not a serious problem in our work. There are differences in many blocks, such as blocks 07, 08, 15, 18 and 20. The reasons may be complex. It could imply some distortion of the MW disc, but our theory is based on a symmetric potential and simple DF model. Sub-structures and bulk motion can also change the shape of the l.o.s velocity DF. 

Finally, the axisymmetry assumption in the torus model may be over-simplified. In fact, many observations indicate that neither the disc star counts nor the disc stellar kinematics are axisymmetric. \citet{2012ApJ...750L..41W} and \citet{2015ApJ...801..105X} claimed that the stellar density of the disc shows quite complicated vertical oscillations in the solar neighborhood and the disc outskirts, respectively. Meanwhile, ~\citet{2013MNRAS.436..101W} and~\citet{2013ApJ...777L...5C} found that the stars have bulk motions in both the radial and vertical directions in the solar vicinity. Also, \citet{2017ApJ...835L..18L} demonstrated that the azimuthal velocity also contains different asymmetric components for stars with different ages within 600\,pc around the Sun. Moreover, \citet{2016arXiv160306262T} identified asymmetric radial and azimuthal velocities from $R\sim8.5$ to 14\,kpc. These peculiar velocities located across the radial range of the outer disc may potentially affect the comparison between the oversimplified torus model and the observed data. Indeed, Fig.~\ref{fig:df8} does show that the observed l.o.s. velocity distributions are slightly asymmetric. However, the asymmetry in these distributions is mostly less than 10\,km\ s$^{-1}$, which should not substantially affect fitting a model since the velocity dispersion values are much larger than the asymmetric velocity by a factor of a few $(<10)$.

To summarise, we have built a model of the Milky Ways's outer disc but it shows that a more sophisticated distribution function is required to model the observed velocity distribution.

\section*{Acknowledgements}
We thank James Binney and the anonymous referee for their very useful comments and corrections. We acknowledge support from the National Natural Science Foundation of China (Grant No. 11333003, 11373032, 11390372, 11403035), and from the 973 Program (No. 2014CB845700). This work is also supported by the Strategic Priority Research Program ``The Emergence of Cosmological Structures" of the Chinese Academy of Sciences (Grant No. XDB09000000). The Guoshoujing Telescope (the Large Sky Area Multi-Object Fiber Spectroscopic Telescope LAMOST) is a National Major Scientific Project built by the Chinese Academy of Sciences. Funding for the project has been provided by the National Development and Reform Commission. LAMOST is operated and managed by the National Astronomical Observatories, Chinese Academy of Sciences.

\bibliographystyle{mnras}

\appendix

\section{The uncertainty of distance}
\begin{figure}
\centering
\includegraphics[width=0.48\textwidth]{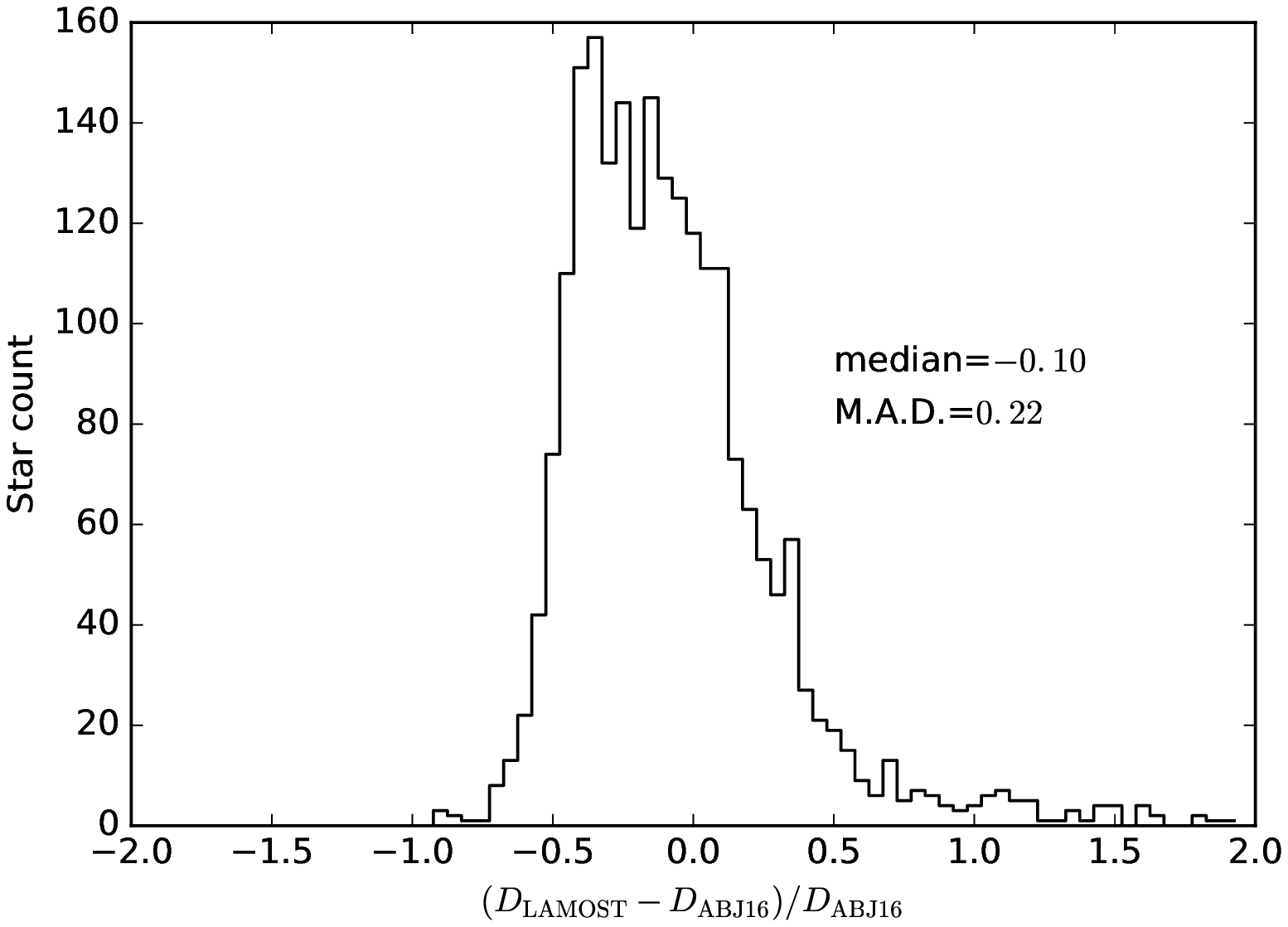}
\caption{The black line shows the relative deviation between the distance $D_{\rm LAMOST}$, which is derived by  \citet{2015AJ....150....4C}  and $D_{\rm ABJ16}$, which is from ABJ16. The y-axis is the count of stars.}
\label{fig:dist}
\end{figure}

The distance uncertainty in \citet{2015AJ....150....4C}  is propagated from the uncertainties in the effective temperature, surface gravity, and metallicity, which are provided by the LAMOST pipeline. However, we find that these uncertainties of the stellar parameters in the LAMOST catalogue are significantly overestimated (also see \citet{2017arXiv170401333S}). To obtain a more realistic assessment of distance uncertainty, we compare the distance for the K giant stars from Carlin et al. with that from \citet{2016ApJ...833..119A} (hereafter ABJ16), who gave the Bayesian derived distance from the TGAS parallax \citep{2016A&A...595A...2G}. We obtain about 2000 common K giant stars with ABJ16 distance smaller than 1 kpc and with relative uncertainty of ABJ16 distance smaller than 20\%.  Figure~\ref{fig:dist} shows the distribution of the relative deviation of the two distances, $(D_{\rm LAMOST}-D_{\rm ABJ16})/D_{\rm ABJ16}$. It can be seen that the Carlin et al. derived distance ($D_{\rm LAMOST}$) is systematically smaller by 10\%. Meanwhile, the dispersion of the relative deviation, measured with M.A.D. (median absolute deviation), is 22\%, which is a factor of 2 smaller than that claimed by \citet{2015AJ....150....4C}. 

\bsp	
\label{lastpage}
\end{document}